\documentclass[12pt,preprint]{aastex}

\newcommand{\rns}{R_{\rm ns}}

\newcommand{\rlc}{R_{\rm lc}}
\newcommand{\zobs}{\zeta}
\newcommand{\zp}{\zeta^\prime}

\newcommand{\mum}{\vec \mu}
\newcommand{\betmin}{\beta}
\newcommand{\rem}{r}
\newcommand{\hem}{h_{\rm em}}
\newcommand{\remp}{r^\prime}
\newcommand{\ro}{\rho}

\newcommand{\thfoot}{\theta_{\rm fp}}
\newcommand{\thgeo}{\theta_{\rm fp,geo}}

\newcommand{\thfootp}{\theta_{\rm fp}^\prime}

\newcommand{\thpc}{\theta_{\rm pc}}

\newcommand{\phl}{\phi_{\rm l}}
\newcommand{\pht}{\phi_{\rm t}}
\newcommand{\phc}{\phi_{\rm f}}
\newcommand{\om}{(\vec \Omega, \mum)}
\newcommand{\dph}{\Delta\phi}
\newcommand{\dphobs}{\Delta\phi_{\rm obs}}
\newcommand{\dphab}{\Delta\phi_{\rm ab}}
\newcommand{\dphret}{\Delta\phi_{\rm ret}}
\newcommand{\etab}{\eta_{\rm ab}}
\newcommand{\rgeo}{r_{\rm geo}}

\newcommand{\rdel}{r_{\rm delay}}

\newcommand{\redge}{r_{\rm edge}}
\newcommand{\rcore}{r_{\rm cr}}
\newcommand{\rcone}{r_{\rm em}}

\newcommand{\wli}{W_0}

\newcommand{\betcor}{\beta_{\rm rot}}
\newcommand{\vbet}{\vec \beta_{\rm rot}}

\newcommand{\vkp}{\vec k^\prime}
\newcommand{\vk}{\vec k}

\shorttitle{Geometry of Radio Emission from Pulsars}
\shortauthors{Dyks, Rudak, Harding}

\begin{document}

\title{On the Methods of Determining the Radio Emission Geometry\\
in Pulsar Magnetospheres}

\author{J. Dyks\altaffilmark{1}}
\affil{Laboratory for High Energy Astrophysics, 
       NASA/GSFC,
    Greenbelt, MD 20771, USA}
\email{jinx@milkyway.gsfc.nasa.gov}

\author{B. Rudak}
\affil{Nicolaus Copernicus Astronomical Center, 87-100 Toru{\'n}, Poland}
\email{bronek@ncac.torun.pl}

\and

\author{Alice K. Harding}
\affil{Laboratory for High Energy Astrophysics, 
       NASA/GSFC,
    Greenbelt, MD 20771, USA}
\email{harding@twinkie.gsfc.nasa.gov}

\altaffiltext{1}{NAS-NRC Research Associate, 
on leave from Nicolaus Copernicus Astronomical Center,
Toru{\'n}, Poland}

\begin{abstract}
We present a modification of the relativistic phase shift
method of determining the radio emission geometry from pulsar
magnetospheres proposed by Gangadhara \& Gupta (2001).
Our modification provides a method of determining radio emission altitudes
which does not depend on the viewing geometry and does not require
polarization measurements.
We suggest application of the method to the outer edges 
of averaged radio pulse profiles
to identify magnetic field lines associated with the edges 
of the pulse and, thereby, to 
test the geometric method based on the measurement of the pulse 
width at the lowest intensity level.
We show that another relativistic method proposed by
Blaskiewicz et al.~(1991) provides upper limits for emission altitudes
associated with the outer edges of pulse profiles. A comparison of
these limits with the altitudes determined with the geometric method
may be used to probe the importance of
rotational distortions of
magnetic field and refraction effects in the pulsar magnetosphere.
We provide a comprehensive discussion of
the assumptions used in the relativistic methods.
\end{abstract}


\section{Introduction}

Since the discovery of pulsars (Hewish et al.~1968) 
the geometry of radio emission
from pulsar magnetospheres was interpreted in terms of emission from 
purely dipolar magnetic fields (eg.~Radhakrishnan \& Cooke 1969; Cordes 1978;
Lyne \& Manchester 1988; Blaskiewicz et al.~1991; Rankin 1993; 
Gil \& Kijak 1993).
This assumption is justified by relatively high emission altitudes
in comparison with $\rns$
inferred for the radio emission ($\sim 0.01\rlc$, where $\rlc=cP/2\pi$ 
is the light cylinder radius and $P$ is the rotation period of
a neutron star with radius $\rns$).
Most importantly, however, radio emission from the dipolar magnetic 
field hopefully can be described by a sufficiently small number of
parameters and a limited number of observational parameters.
The word ``hopefully" reflects a second crucial assumption
applied for the radio emission beam:
that distinguishable features in pulse profiles
(conal components, the outer edges of a profile) are associated with
the beam structure (eg.~of concentric hollow cones of enhanced radio emission)  
which in the reference frame
corotating with the neutron star (CF)
is symmetric 
with respect to the plane containing 
the dipole magnetic moment $\mum$ and the rotation axis.
Without this disputable assumption, the number of parameters
required to determine the emission geometry increases
considerably
and one is left with a multi-parameter theory to be compared
with data from which only a few parameters can be deduced.
Hereafter, the axial symmetry (in CF) of the radio emission beam
is assumed; the problem of its justification is presented in Section
4.1.

To further constrain the parameter space,
these two assumptions (I -- dipolar magnetic field; II -- 
symmetry of radio beam) 
are often supplemented with two additional simplifications: III -- it is
assumed that in the CF the bulk of radio waves 
is emitted in the direction which
is tangent to the local magnetic field at an emission point;
IV -- identifiable features in pulse profiles
(eg.~maxima of conal components) observed in a narrow frequency band
are interpreted as radiation from a very narrow range of altitudes.
The assumption No.~III requires relatively large Lorentz factors
$\gamma$ of radio emitting electrons. Hereafter we assume that 
the broadening of the pulse features due to the
radiation from the low energy electrons is negligible, although it may
be significant in reality. Our assumption is fully justified in the case
of the most popular radiation mechanism -- the curvature radiation,
because the electron Lorentz factors must exceed $\sim 100$ for the
curvature spectrum to extend up to the observation frequency ($\sim 300$
MHz). Curvature radiation from low energy electrons ($\gamma \sim 10$)
cannot broaden the profile because it does not extend to $10^2$ MHz. 
The model-dependent 
estimates of $\gamma$ cover a large range between $50$ and $10^4$
(eg.~Ruderman \& Sutherland 1975; Machabeli et al.~2001; Melrose 2004).
For the rather conservative estimate of
$\gamma \sim 100$ the corresponding angular size of the single
electron radiation beam is $\gamma^{-1}\sim 0.5^\circ$, 
which is an order of magnitude smaller than the typical width of the radio
beam corresponding to the mean pulse profile.

With the assumptions I--IV,
the radio emission geometry becomes completely
determined by four parameters: 
$\alpha$, $\betmin$, $\rem$, and $\ro$.
Their meaning is the following:
$\alpha$ is an inclination of the magnetic dipole
with respect to the rotation axis (ie.~the angle between the magnetic
moment $\mum$ and the angular velocity $\vec\Omega=2\pi P^{-1} \hat z$).
$\betmin$ is an ``impact angle", 
ie.~the minimum angle between an
observer's line of sight and $\mum$. This parameter is often replaced
by $\zeta=\alpha+\betmin$ -- 
the angle between the observer's line of sight 
and the rotational axis.
The next parameter, $\rem$, is the radial distance of the radio emission
region measured from the center of the neutron star. 
Hereafter, we will
often use its normalized value $\remp=\rem/\rlc$.
The last parameter, $\ro$, is the half opening angle of the
radio emission
cone/beam, ie. it is the angle between the direction of radio emission
in CF and $\mum$. 
These four parameters determine completely the emission geometry in the
sense that any additional information about the emission region
(eg.~coordinates of emission point in the dipole frame, with z-axis
along $\mum$) can be easily derived from textbook
equations for the dipole geometry and spherical trigonometry
(eg.~Radhakrishnan \& Cooke 1969; Cordes 1978; Gil et al.~1984).

One desired additional piece of information
is an answer to the question ``which magnetic field lines
does the observed emission come from?".
Following standard conventions, we identify magnetic field lines by
$\thfoot$ -- the colatitude of foot points of the magnetic field lines
at the neutron star surface.   
With the radio emission altitude $\rem$ and the beam radius $\ro$ 
determined, the
value of $\thfoot$ can be easily calculated. 
The parameter $\thfoot$ is often expressed in terms of a fraction of
the polar cap angle: $\thfootp=\thfoot/\thpc$, where 
$\thpc=\arcsin((\rns/\rlc)^{1/2})$.

These four quantities are to be deduced from radio data. 
In most cases, however, a model-dependent analysis of the radio data
at a given frequency provides us with only two quantities:
$\betmin$, and $W(f)$, where $W(f)$ is the
apparent width of the radio pulse profile, usually defined by some simple
criterion (eg.~measured at some fraction $f$ of maximum intensity; a wide
variety of $f$ is employed: $f=0.0005$ (Kijak \& Gil 2003), $f=0.1$
(Blaskiewicz et al.~1991), $f=0.5$ (Rankin 1993), $f=1$ (Gangadhara \&
Gupta 2001)). 
The value of the inclination angle $\alpha$ could in principle be
determined along with $\betmin$ from the ``rotating vector"
 model of polarization position angle swings
(Radhakrishnan \& Cooke 1969). In practice, however, a fit of the model
to the observed position angle curve is much less sensitive to $\alpha$ than to
$\beta$ (Rankin 1993). 
Only in exceptional cases can both of these parameters be derived
(Lyne \& Manchester 1988; Blaskiewicz et al.~1991;
von Hoensbroech \& Xilouris 1997).
This led Rankin (1990) to a formulation of a method
of determining $\alpha$ which assumes that the apparent width of 
the core component does not depend on the impact angle
$\betmin$ and that in all pulsars the core component originates from
the same, low altitude. 
The observed width of the core component 
becomes then a universal function of
the pulsar rotation period $P$ and the dipole inclination $\alpha$,
allowing, thus, to determine the latter parameter.

We are provided, therefore, with 
just three parameters ($\alpha$, $\betmin$, $W$) instead of four.
This limitation prompted the development of 
two kinds of methods to determine the geometry of the radio
emitting region: 1) A purely geometric method
which assumes that the lowest detectable emission at the leading and at
the trailing edge of a radio pulse originates at the last open magnetic
field lines, with $\thfoot=\thpc$. With the fourth parameter assumed
{\it a priori},
the method makes it possible to determine the radial position 
of radio emission
$\rem$ from the observed pulse width $\wli$ 
(eg.~Cordes 1978; Gil \& Kijak 1993; 
Kijak \& Gil 1997; 1998; 2003; Kijak 2001; hereafter we will use
the index `0' to refer to the pulse with $W(f)$ at the lowest intensity
level, ie.~practically at $f=0.0005 - 0.1$). 
The derived altitudes are a few tens of $\rns$ at observation frequency
$\nu\sim1$ GHz. 
2) Methods of the second kind are able to derive 
the fourth observational parameter by a measurement of phase shifts
of some profile features with respect to some fiducial points.
The methods are relativistic in the sense that the phase shifts are
caused by combined effects of aberration and propagation time delays due
to the finite speed of light $c$ (for brevity, the latter effect
will hereafter be called retardation). The second methods are superior
to the
geometric method in that they do not assume {\it a priori} the value
of the fourth parameter. However, they must rely on additional assumptions
about the radio pulse profile. Gangadhara \& Gupta (2001, hereafter
GG2001) measure the
relativistic phase shift of conal components 
with respect to the core component, which is
assumed to originate from much lower altitude than the cones which surround it.
Blaskiewicz et al.~(1991, hereafter BCW91) 
measure the shift of pulse edges
with respect to the center (or the ``inflection point") 
of the position angle curve which is assumed to 
originate from the same altitude as the emission at
the outer wings of the profile.
On average, the method of Blaskiewicz et al.~(1991) (hereafter BCW method)
gives emission radii in rough
agreement with that of the geometric method
(Gil \& Kijak 1993; Kijak \& Gil 1997; 1998; 2003). 
However, the method of GG2001 predicts
notably larger values of $\rem$ (by about factor of 2) 
than the geometric method.
Unlike the geometric method (Cordes 1978; Gil \& Kijak 1993),
both relativistic methods (of GG2001 as well as BCW91) \emph{in principle} 
make it possible to identify the radio
emitting field lines. GG2001 and Gupta \& Gangadhara (2003, hereafter
GG2003) find $\thfootp$ in the range $0.22-0.74$ for radio emission 
observed at maxima of conal components.
Unfortunately, the method of GG2001 can be applied only for pulsars
having unambigously identifiable pairs of  
conal components (and which possess a core
component). So far, this has required application of
 a window thresholding technique
(hereafter WT technique, GG2001) and limited application of the
method only to a handful of the brightest objects.
On the other hand, the BCW method suffers from a difficulty in finding the
center of the position angle curve, and, as we show below, it yields
values of $\thfootp$ exceeding 1.
For convenience, hereafter we refer to the relativistic methods of GG2001 
and BCW91 with the term
``relativistic phase shift methods" (RPS methods).

In Section 2 we revise the method of GG2001,
which results in a new formula for radio emission
altitudes and furnishes the method with new interesting features.
As noted in GG2003, the original method of GG2001 yielded
radio emission altitudes larger than those derived with the 
geometric method. Our revision removes part of this discrepancy.
In Section 3 we propose to apply the method to
the outer edges of averaged
pulse profiles. This will hopefully provide 
a test of the main assumption of the
geometric method about the value of $\thfootp = 1$ for the beam edge.
As an example, we try to perform such a test using 
the method of BCW91.
In Section 4 we discuss in detail the assumptions 
of the relativistic methods.
Our conclusions are in Section 5.

\section{The modified relativistic phase shift method}
\label{correction}

The RPS method of Gangadhara \& Gupta (2001)  
applies to pulsars with both core and cone emission, 
ie.~for M and T pulsars in the classification scheme of Rankin
(1983). 
Questionability of the conal pattern of radio emission beam will be
discussed in Section \ref{asymmetry}.
The model assumes that
in the reference frame corotating with the neutron star
the cones are symmetric around the narrow core beam.
Obviously, in the case of the dipolar magnetosphere
the second assumption can be most naturally accounted for by
a model of the radio beam consisting of a narrow core beam centered 
at the dipole axis and a few nested hollow cones of enhanced
radio emission surrounding the core. 
Magnetic field lines corresponding to a given cone have the same
$\thfoot$ and the same half-opening angle $\ro$ at a fixed altitude. 
In fact, the second assumption can be made less stringent: it is sufficient
for the conal beams to be symmetric with respect to the plane containing
the core component and the rotation axis. Thus, the ``conal" beams may
have eg.~elliptical crossection with the core beam located in the plane
containing the ellipse center and the rotation axis (ie.~not necessary
at the ellipse center). 
The possibility of the elliptical beam is discussed eg.~in Mitra \& Deshpande
(1999).

The method starts with identifying the core component in the radio pulse
profile. Then one attempts to identify the same number of conal components 
on both
the leading and the trailing side of the core component.
The innermost pair of two conal components
located on both sides of the core
is identified as emission from the same, innermost
hollow cone of the radio beam. The next pair of conal components which
bracket/flank the innermost pair is interpreted as emission from the
next-to-innermost hollow cone of the radio beam, and so on.
For each conal pair the observed phase of the leading component ($\phl$)
and the phase of the trailing component ($\pht$) is measured with
respect to the core component which defines the phase $\phi=0$.
As emphasized by GG2003 in all cases with clear identification of cones,
the conal components were shifted toward earlier phases with respect to the 
core component, ie.~$|\phl| > \pht$.
This effect was noticed already by Gil (1985), who proposed the
retardation effect to explain the asymmetry.
More correctly,
GG2001 interpreted the asymmetry as a forward shift
of conal components due to combined effects of aberration of conal
emission and retardation of core emission. 
The (negative) relativistic phase shift $\dphobs$
between the midway point of conal pairs and the core component
is equal to:
\begin{equation}
\dphobs = \frac{\pht - |\phl|}{2}
\label{ph}
\end{equation} 
and the separation between the maxima of conal components
is given by
\begin{equation}
W = \pht + |\phl|
\label{w0}
\end{equation}
The observed positions of the two conal components  
($\phl$ and $\pht$)
measured with respect to the core component are 
the \emph{two} most desired observational parameters which replace the
single $W$ parameter. 
Beyond question, recognizing this point is a great insight of
Gangadhara \& Gupta (2001).

The next step in the method of GG2001 was to express $\dphobs$ in terms
of radial distance of conal radio emission $\rcone$.
In the absence of the aberration and retardation effects (hereafter AR
effects), the
center (ie.~the maximum) of the core component as well as the center of
the conal pair would be observed at the same fiducial 
phase $\phc$ which formally
corresponds to emission from $r=0$.
In general, the radial distance $\rcone$ of the conal emission is different
from the radial distance $\rcore$ of the core emission.
Therefore, the AR effects shift
the center of the conal pair by $\dph(\rcone)$ with respect
to $\phc$, whereas the center of the core component is shifted with
respect to $\phc$ by a
different amount of $\dph(\rcore)$. The observed shift is the difference
between these: $\dphobs = \dph(\rcone) - \dph(\rcore)$.

The value of $\dph$ is a sum of a shift due to the aberration
and a shift due to the retardation: $\dph = \dphab(\rem) + \dphret(\rem)$,
where the radial distance of radio emission $\rem$ may
refer either to $\rcone$ or $\rcore$.
The retardation shift is equal to
\begin{equation}
\dphret \simeq -\frac{\rem}{\rlc},
\end{equation}
regardless of $\alpha$ and $\betmin$,
at least as long as the lowest order ($\sim \remp$) relativistic
effects are considered.

To first order in $\rem/\rlc$ (cf.~Appendix \ref{appen}; GG2001),
the aberration changes the direction of photon emission by an angle:
\begin{equation}
\etab \simeq \frac{v_{\rm rot}}{c} \simeq \frac{\rem}{\rlc}\sin\zeta
\label{etab}
\end{equation}
where $\vec v_{\rm rot}=\vec \Omega \times \vec \rem$ 
is the local corotation velocity at the emission
point. 
Following the already published work (Cordes 1978; Phillips 1992), 
GG2001 assumed that the resulting
phase shift $\dphab$ is equal to $-\etab$.
This is, however, not true in general, because
the aberrated emission direction, 
when rotated by
$360^\circ$ around $\vec \Omega$, does not delineate a great circle on a sphere centered at the star.
For this reason, the aberrational shift should be written in the form:
\begin{equation}
\dphab \simeq -\frac{\etab}{\sin\zeta} \simeq -\frac{\rem}{\rlc}.
\label{aber}
\end{equation}
Eq.~(\ref{aber}) reflects the fact that any decrease in the aberration
angle $\etab$ caused by the smaller
corotation velocity near the rotation axis is cancelled out by the small
circle effect (eq.~\ref{obv}). A rigorous derivation of this approximate
result can be found in Appendix \ref{appen}.
Thus, the total relativistic shift with respect to $\phc$ is equal to
$\dph \simeq -2\remp$, whereas the observed shift of the conal pair 
with respect to the core is equal to $\dphobs = -2(\rcone -
\rcore)\rlc^{-1}$.
The resulting formula for the altitude of the conal emission reads
\begin{equation}
\hem = \rcone - \rcore \simeq -\frac{\dphobs}{2}\rlc.
\label{alti}
\end{equation}
In the case $\rcore \ll \rcone$, considered by GG2001:
\begin{equation}
\rcone \simeq -\frac{\dphobs}{2}\rlc.
\label{radius}
\end{equation}
This formula gives emission radii smaller by a factor
$a\simeq (1 + \sin\zeta)/2$ than those obtained by GG2001
(cf.~their eq.~9).
As shown in Appendix \ref{appen}, 
the approximation (\ref{radius}) 
holds with accuracy of $\sim
10$\% for $\remp \la 0.01$.
For objects with moderate or large dipole inclinations ($\alpha \simeq
40^\circ - 90^\circ$) this is a minor correction, 
however, for nearly
aligned rotators the radii become almost two times smaller.
The method of GG2001 applies to broad pulse profiles 
with well resolved conal components.
This criterion favours objects with small $\alpha$, and, therefore,
makes the modification important.
GG2003 find particularly large emission radii for two objects, one of
which (B2111$+$46) has very small inclination angle $\alpha=9^\circ$.
For this object, viewed at the angle
$\zeta=11.4^\circ$, our modification results in emission radii which are
smaller by a factor of $0.6$ than those derived by GG2003.
For the second problematic pulsar (B2045$-$16), viewed at a considerably
larger angle $\zeta=37.1^\circ$, 
our modification does not reduce the
radii significantly. GG2003 report some difficulties with unambiguous
identification of conal pairs for this pulsar.\\
Using the values of $\phl$ and $\pht$ published 
by GG2001 and GG2003 we recalculated the radio emission altitudes
for objects studied therein. The resulting
values are given in Table 1 (columns 5 and 6). 
The errors of $\hem$ ($\sim 22$\%) are much larger
than the errors of $\phl$ and
$\pht$ ($\sim 3$\% as measured by Gangadhara \& Gupta), because $\hem
\propto \dphobs$ and $\dphobs$ is a very small difference between $\pht$
and $|\phl|$ (eq.~\ref{ph}).

We emphasize that our modification is not barely a rescaling
of the method of GG2001, but it furnishes their method with completely
new features. 
Unlike the original equation (9) of GG2001,
our formula for the emission radius (eq.~\ref{radius})
\emph{does not} depend on $\alpha$ and $\betmin$, which is a valuable
feature given the problems with determining $\alpha$.
A method of BCW91 is also independent of $\alpha$
and $\betmin$, however, it requires high-quality 
polarization measurements, and can
only be applied to pulsars with well ordered position angle swings.
Our modification of the method of GG2001 provides
a method for determining $\rem$, 
which not only can do without the knowledge of $\alpha$ 
and $\betmin$,
but at the same time 
it does not require the polarization measurements.
It is based solely on information contained in M, and T-type profiles.
The applicability of different methods to different
sets of available data is summarized in Table \ref{appl}.
Since the presence of a core in pulse profiles is often
accompanied by a disordered shape of the position angle curve 
(eg.~Rankin 1993),
the two methods (ie.~the delay-radius relation of BCW91 
and our eq.~\ref{radius}) 
can complement each other.
On the other hand, this is a disadvantage, because it makes it difficult
to compare the methods. 

To identify magnetic field lines from which the conal emission originates,
it is necessary to determine the half opening angle $\ro$ of the emission
cones. This is accomplished with the help of the cosine formula of
spherical trigonometry, which connects the values of $\ro$, $\alpha$,
and $\zeta$ with the observed separation of conal components
$W(f=1)$:
\begin{equation}
\cos\ro = \cos\left(\frac{W}{2}\right)\sin\alpha\sin\zeta + \cos\alpha\cos\zeta.
\label{ro}
\end{equation}  
This time the knowledge of $\alpha$ and $\zeta$ is necessary.
Since the formula for $\ro$ does not depend on $\rem$ the values
of $\ro$
calculated by GG2001 and GG2003 are correct.
  
With $\ro$ and $\rem$ determined from eqs.~(\ref{ro}) and
(\ref{radius}), one can calculate the colatitude 
of footprints
of active magnetic field lines at the star surface
with the dipolar formula:
\begin{equation}
\thfoot = \arcsin\left[\left(\frac{\rns}{\rem}\left(
\frac{2}{3} - \frac{1}{6}\left(x + \left(x^2 +
8x\right)^{1/2}\right)\right)\right)^{1/2}\right],
\label{thfoot}
\end{equation}
where $x=\cos^2\ro$ (see Appendix \ref{appen}). 
In the small angle approximation ($\ro \ll1$)
the formula reduces to the well known approximate form:
\begin{equation}
\thfoot \simeq \frac{2}{3}\ro\left(\frac{\rns}{\rem}\right)^{1/2}.
\label{thfootapp}
\end{equation}
Obviously, since the revised method predicts lower emission radii
than its original version but the same opening angles $\ro$,
it must yield larger footprint colatitudes $\thfoot$.
Since, $\thfoot \propto \rem^{-1/2}$ the values of $\thfoot$ increase
only by a factor of $b=(2/(1 + \sin\zeta))^{1/2}$ between $1$ and $1.4$.
Consequently, the modified values of $\thfootp=\thfoot/\thpc$ 
for objects studied by GG2001
and GG2003 are given in the last column of Table 1.
They range from 0.28 to 0.88. The values of $\thfootp$ for the outermost
cones cover the range between 0.35 and 0.88.
The errors of $\thfootp$ are based only on 
the errors of $\phl$ and $\pht$
determined by GG2001 and do not include the large 
uncertainties of $\alpha$, and
$\betmin$. 

Since our revision rescales the values of $\rem$ and of $\thfoot$
by factors which are the same for a given pulsar, the main findings 
of GG2001 and GG2003 (ie.~higher altitudes for broader cones
and lower altitudes for higher radio frequencies) remain valid.

\section{Comparison of the RPS methods with the geometric method}
\label{comparison}

In addition to the conal components, there is one more feature in radio
pulse profiles which for a long time has been believed 
to be symmetric with respect to the $\om$ plane 
-- the outer edge of the radio beam.
The assumption about the symmetry 
has always been present in the traditional, geometric method of
determining radio emission altitudes (eg.~Kijak \& Gil 1997).
Therefore,
the method of GG2001 may also be applied to the
edges of radio beam, not only to maxima of the conal components.
Hereafter, the RPS method of GG2001 with this assumption
will be referred to as the ``outer edge
relativistic phase shift method" (OERPS method). The method requires a
measurement of $\phl$ and $\pht$ for the outer edges of an
\emph{averaged} pulse profile. Then calculations of the original
method of GG2001 can be performed (equations \ref{ph} -- \ref{thfoot}) 
to determine
both the altitude and the locations 
$\thfootp$ of magnetic field lines for the radio
emission at the pulse edge.
Thus, the OERPS method may be used to test the geometric method 
which a priori assumes
$\thfootp({\rm edge}) = 1$. 

The OERPS method does not require an identification of conal
components within the pulse profile. 
This method will work without the analysis of single pulses 
(with the WT technique)
and it can
be applied directly to averaged pulse profiles with large signal to
noise ratio. 
A significant limitation of the OERPS method is that it requires
very accurate measurements
of the observed positons $\phl$ and $\pht$ of the leading 
and the trailing edge of the radio pulse profile.
These are much less defined than the positions of conal maxima.
Therefore, the method may give less accurate results than the
original RPS method of GG2001, 
unless very precise determination of $\phl$ and
$\pht$ is achieved.
GG2001 and GG2003 found that shifts of outermost cones' maxima 
with respect to the
core may in some cases be as large as $3^\circ$,
and they were able to measure some shifts with accuracy of $\sim
0.05^\circ$ ie.~a few percent.
A much larger error (typically $\sim 30$ \%) is inherent in 
the OERPS method and in the method of BCW91
which must deal with locating
the ``lowest intensity" points.

Because we have no access to high-quality radio data to apply the OERPS method, 
instead, 
below we try to use the BCW method to identify the active magnetic field
lines.
Like the OERPS method, 
the method of BCW91 provides the value of $\rem$ 
in a way which is independent of $\alpha$, $\betmin$, and, most
importantly, of $\thfoot$. 
The lowest intensity width $\wli$ 
(usually measured by BCW91 at the level of $f=0.02$
and used to determine the center of a profile), 
along with $\alpha$, $\beta$, and $\rem$,
provides all necessary information to calculate $\thfoot$
with the help of equations (\ref{ro}) -- (\ref{thfoot}), ie.~in the same
way as in GG2001.

However, since different components of a pulse profile may originate
from different altitudes, 
the values of $\rem$ derived with some methods
may represent an average over several components.
To calculate $\thfoot$ for the pulse edge, 
it is necessary to know the radial distance
$\redge$, which refers to the outer edges of the pulse profile.
Establishing the relation between the pulse-averaged emission radius 
$\rem$, and the ``edge radius" $\redge$ is an important step in
determining $\thfoot$ for the outer edge of the pulse profile.
Normally, calculating $\thfoot$ for the pulse edge would require
using the equations (\ref{ro}) --
(\ref{thfoot}) with $\rem=\redge$. 
An easier way is to use the equation (\ref{thfootapp}) which implies that
the ratio of $\thfoot$ 
corresponding to $\redge$ determined with some method,
to $\thgeo$ of the geometric method is:
\begin{equation}
\frac{\thfoot}{\thgeo} \simeq \thfootp \simeq 
\left(\frac{\rgeo}{\redge}\right)^{1/2}, 
\label{ratio}
\end{equation}
because $\thgeo = \thpc$ by assumption, which we want to verify.
In the above formula $\thgeo$ and $\rgeo$ refer to 
the geometric method,
whereas $\thfoot$ and $\redge$ refer to another method.\footnote{For
brevity, in eq.~(\ref{ratio}) we skip the index `edge' at $\thfoot$ 
and $\thfootp$, which
hereafter should be understood as referring to the outer edge
of the pulse profile.} 
The above equation holds only when
$\redge$ (and $\thfoot$) refer to the outer edge of a pulse
profile, just as the values of $\rgeo$ do.
Emission radii provided by the BCW method (hereafter denoted by $\rdel$)
are calculated under the assumption that all components in a pulse profile
originate from the same altitude, ie.~$\redge = \rdel$ by assumption.
Thus, inserting $\redge=\rdel$ into eq.~(\ref{ratio}) one can calculate
$\thfootp$ predicted by the BCW method for the pulse edge.

We use the values of $\rdel$ and $\rgeo$ calculated for the same data
set by BCW91 and presented in their table 3.
The values of $\thfootp$ derived with the BCW method 
for seven clean-cut cases 
with reasonably small statistical errors, and most probably free from
systematic errors,
are presented in column 3 of Table 2. Column 4 presents a range of
$\thfootp$ allowed by 1 $\sigma$ errors derived for $\rdel$ and $\rgeo$
by BCW91 (table 3 therein). The last column shows the level of
consistency of the two methods (in $\sigma$), ie.~the level at which the
derived values are consistent with $\thfootp = 1$.

According to Table 2, 
for four cases (B0301$+$19 and B0525$+$21, both at
0.43 and 1.4 GHz) the BCW method yields $\thfootp > 1$, 
ie.~radio emission
from the closed magnetic field line region. The inconsistency with 
$\thfootp = 1$ is at $\sim 2\sigma$ level.
The result $\thfootp > 1$ is equivalent to $\redge < \rgeo$
(cf.~eq.~\ref{ratio}), and, therefore may result from
underestimating $\redge$ and/or from overestimating $\rgeo$.
As noted by BCW91, contrary to the assumption $\redge=\rdel$,
the values of $\redge$ may actually be larger than $\rdel$, 
because $\rdel$
represents an average over the pulse profile, and inner parts of the
profile may originate from lower altitudes than the outer 
edges of the pulse (eg.~Rankin 1993; GG2001).
Below we show, however, that even if 
the central parts of the pulse profiles originated from
the star surface, the values of $\redge$ cannot exceed $2\rdel$.
In the BCW method, the values of $\rdel$ are
derived from the relative position of the outer edges of the profile
and the center of
the position angle (PA) curve.
Let us assume that the center of the PA curve is 
determined mostly by the central parts of
profiles which may originate from lower altitudes
than the beam edge.
Let us define a fiducial phase zero as a moment at which an observer
detects a light signal emitted from the neutron 
star center when the dipole axis
was located in the plane containing the rotation axis and the observer's
position.
The star's rotation shifts the profile center (determined by the
midpoint between the \emph{outer edges} of the pulse) toward earlier
phases
by $-2r_{\rm edge}/\rlc$ with respect to the fiducial phase, 
where $r_{\rm edge}$ is the radial distance
for emission at the edge of the pulse profile. 
At the same time, the rotation delays
the center of the position angle swing by $2r_{\rm in}/\rlc$
with respect to the fiducial phase, where
$r_{\rm in}$ refers to the radiation from inner parts of the profile
which determine the position of the center of the PA swing, according to
our working hypothesis.
For the case considered by BCW91 (ie.~$r_{\rm edge} = r_{\rm in}$), 
the total shift
is $4r_{\rm edge}/\rlc$. 
For the other limiting case $r_{\rm edge} \gg r_{\rm in}$
the total shift is two times smaller: $2\redge/\rlc$, since the delay
of the center of the PA curve with respect to the fiducial phase is
negligible.
In this extreme case the value of $\redge$ 
derived from the shift $\Delta\phi_{\rm
PA}$ between the center of the PA curve and the profile center
is given by:
\begin{equation} 
\redge = \frac{\Delta\phi_{\rm PA}}{2}\rlc = 2\rdel\ \ \ \ ({\rm for}\ \redge
\gg r_{\rm in}).
\label{redge}
\end{equation} 
Therefore, in general the emission radii for the beam edge
may be at most two times larger than $\rdel$ derived by BCW91 for the case of 
$r_{\rm edge} = r_{\rm in}$:
\begin{equation}
\redge \le 2\rdel,
\label{limit}
\end{equation} 
Using the upper limit of $\redge = 2\rdel$ in eq.~(\ref{ratio}), 
would decrease the values of $\thfootp$ in Table 2
by a factor of $2^{-1/2}$.
In spite of the large error range for $\thfootp$
(column 4 in Table 2),
this would make only \emph{two} of the four problematic
values consistent with 1 within the level of $1\sigma$.
The other two values of $\thfootp$ 
(for B0301$+$19 at 1.42 GHz and B0525$+$21 at 0.43 GHz) 
would still remain inconsistent with 1 at $1\sigma$ level.
Thus, lower emission altitudes for the 
central parts of profiles are not able to 
remove the disagreement completely.
Indeed, different locations
of emission regions for the conal peaks and for the bridge between them
have been proposed for the two pulsars based on different spectra
and fluctuation properties (Backer 1973; Rankin 1983; BCW91).
For B0525$+$21, BCW91 suggest the contribution of 
a quadrupole component of stellar magnetic field
as a source of the error in $\rdel$. 
The underestimate of $\rdel$ may also be caused by the
rotationally-induced magnetic field line sweep-back, since the effect
produces a shift of opposite sign to the shift $\Delta\phi_{\rm PA}$ found
by BCW91. 
The significance of this effect will be investigated in a forthcoming
paper.

The result $\thfoot > 1$ (or $\redge < \rgeo$) could also arise due to
an overestimate of $\rgeo$.
This may be caused by refraction effects,
which would broaden the radio beam with altitude (Lyubarskii \& Petrova
1998). However, this effect could be compensated (or even dominated)
by the unavoidable underestimate of $\rgeo$ which is
inherent to the method of
its derivation: the geometric method associates the maximum value of
$\thfootp=1$ with the observed pulse width, whereas the true value 
of $\thfootp$ may well be
smaller due to an inactivity of the outer parts of the polar cap
and/or due to a limited sensitivity threshold.
Therefore, in the case of the rectilinear propagation of radio waves,
the values of $\rgeo$ should be considered as lower limits
for $\redge$.
In consequence, it is worth noting that BCW91 underestimated
their geometric radii $\rgeo$, 
because these were calculated for 
the 10\% intensity level ($f=0.1$) instead of the lowest intensity level,
which BCW91 usually identified with 2\% of maximum intensity.
Therefore, the values of $\thfootp$ in Table 2 are underestimated,
and the actual disagreement of the BCW method
with the condition $\thfootp \le 1$ is even larger than given
in the last column of Table 2.

For B1914$+$13 an upper limit for $\thfootp$ equal to $0.8$ is given in
Table 2 (based on $\rgeo$ and $\rdel$ from table 3 of BCW91).
This is the only case (of single core type -- S$_{\rm t}$),
for which $\thfootp < 1$. 
Obviously, this could equally well imply the 
inactivity of the outer part of the polar cap as our inability
to detect the outer wings of radio profiles.
In either case,
the assumption that $\thfootp=1$ at the \emph{observed}
profile edge would be invalidated. We emphasize, 
that like any value of
$\thfootp$ in Table 2, the upper limit of $\thfootp=0.8$ may be
underestimated because of the relatively high level of intensity (10\%)
for which the values of $\rgeo$ were calculated in BCW91.

Thus, by selecting a sample of pulsars studied by BCW91 with the smallest
statistical errors, and least probable to be affected by
systematic errors, we find that the BCW method 
implies the radio emission from
the region of closed magnetic field lines, (ie.~with $\thfootp > 1$),
at the level $>2\sigma$ in some cases.
The way out would be to postulate either significant
magnetic field deviations from the static dipole shape, 
or refraction effects (Lyubarskii \& Petrova 1998).
The distortions of the magnetic field may be caused by the presence
of the quadrupolar stellar magnetic field (BCW91), 
or due to the rotationally induced
sweep-back (Shitov 1983).

\section{Discussion of assumptions of the relativistic phase shift methods}
\label{weakpoints}

\subsection{Asymmetry of the radio beam}
\label{asymmetry}

To provide reliable estimates of $\rem$
the RPS method of GG2001 requires the symmetry of cones of enhanced
radio emission with respect to the plane containing the core component
and the rotation axis. To calculate $\thfoot$, the core/cones system must be centered
at the dipole axis. The OERPS method requires the same symmetry for the
outer edge of the radio beam, and the BCW method requires 
the symmetry of the beam edge with respect to the $\om$ plane,
regardless of the position of the core component.
All these assumptions are related to the old problem: is the beam shape
conal or patchy? The patchy beam would invalidate all these methods.
Fortunately, however, statistical analyses of distribution of components
within the radio pulse provide arguments for the conal shape (eg.~Mitra
\& Deshpande 1999; Kijak \& Gil 2002; Gil et al.~2002).
It seems that arguments for patchy structure of the beam
(Lyne \& Manchester 1988, hereafter LM88)
are not equally strong. In addition to the effects described by
Gil et al.~(1993), the aberration and the retardation may enhance
the \emph{apparent} patchiness of the beam.
This is because central components of pulse profiles seem to originate from 
lower altitudes than the outer components (eg.~Rankin 1993; GG2001). 
The aberration
and retardation effects shift the relative positions of different components
in phase by $\Delta \phi \simeq 2\Delta \remp$, where $\Delta \remp$ is the
difference of emission radii between different components
in units of $\rlc$.
Estimating $\Delta \remp \sim 0.01$ one obtains $\Delta \phi \sim 1^\circ$
which amounts to $\sim 13$\% of pulse width observed for pulsars
with large dipole inclinations ($\alpha \sim 90^\circ$). 
Given that observed separation between components is much smaller than
the pulse width (say $20-40$\%), 
these shifts, different for different objects because
of diversity of $\rem$, along with other factors (eg.~inexactly
determined 
impact angles $\betmin$)
could easily produce the apparently random
distribution of components within the pulse window, ie.~in
figures like the fig.~12 in LM88. 

Other arguments for the conal beam shape are provided directly by the WT
technique of GG2001. For 6 (out of 7) objects studied by GG2001 
and GG2003,
the same number of conal components was found on both sides of the core
component, and only B2045$-$16 is a \emph{possible} exception.  
In all cases, strong (ie.~easily identifiable) components
on the leading side corresponded to strong components on the trailing
side, whereas new weak components, detected with the WT technique
on one side of the core
were always associated with weak components on the other side.   
It is natural to interpret this coincidence 
in terms of the conal, not patchy beam shape.
Another interesting implication of this relation is the
possibility of applying the method of GG2001 to a larger number of
T and M pulsars, without the neccessity
of identification of all cones with the WT technique.

Thus, the conal beam shape seems to be consistent with observations.
On theoretical grounds, the conal shape is supported by a model of sparks
rotating around the magnetic pole due to the $\vec E\times \vec B$ drift
(Ruderman \& Sutherland 1975; Gil \& Sendyk 2000).
The model, as well as the conal beam hypothesis,
 is supported by observations of subpulse drifts 
(Gil \& Krawczyk 1997;
Deshpande \& Rankin 1999; Vivekanand \& Joshi 1999).

\subsection{The reference altitude}

All three RPS methods rely on a measurement of a shift between three
identifiable points within a pulse profile and make assumptions about
relative altitudes of these points.
The method of GG2001 provides the \emph{altitude} of the conal radio
emission measured with respect to the altitude of the core emission
(eq.~\ref{alti}). To identify the active magnetic field lines 
(ie.~to calculate $\thfootp$) it is necessary either to know $\rcore$
or to rely on the assumption
$\rcore \ll \rcone$ (as we did in Table 1) which effectively means that
the derived values of $\thfootp$ are upper limits. 

There is a very simple geometric argument against the hypothesis
that the core components originate from the very vicinity of the star
surface: With the only exception of B1237$+$25, for all other objects
studied by GG2001 and GG2003, and listed in our Table 1, the impact
angles $\betmin$ are \emph{larger} than the half opening angle 
of the last open field lines at the surface of the neutron star with
radius $\rns=10$ km. Had the central components (identified as cores)
originated close to the surface, our line of sight should miss them.
Therefore, the impact angles determined for these objects from fitting
the position angle swing are in clear disagreement with
the assumptions of the method of determining $\alpha$ proposed by Rankin
(1990). 
Assuming rectilinear propagation of radio waves,
this discrepancy can be interpreted only in two ways:
1) either the radial distance for the core emission is much larger 
than $\rns$, or 2) the impact angles $\betmin$ are systematically
overestimated. 
In the first case, the dipole inclination angles $\alpha$ derived with 
Rankin's method would be underestimated, at least as long as
the width of the core beam is assumed to reflect the width of the open
field line region at a given altitude.
For the time being, it seems to be a matter of personal preference
which of the possibilities (either 1. or 2.) is the case -- Rankin
(1990) assumed that the values of $\betmin$ are in error.
Regardless of the choice, however, either $\alpha$ (based on the
assumption
$\rcore\sim\rns$) or $\betmin$
\emph{must} be in error, since they contradict each other. 
Therefore, the values of $\thfootp$ given in
the last column of our Table 1 should be treated with caution.

PSR B0450$-$18, studied by GG2003
has the largest impact angle 
($\betmin \simeq 4^\circ$) among all the objects listed in Table 1.
The half opening angle of the open field line region at the surface
of the pulsar is equal to $1.5\thpc\simeq1.68^\circ$. 
To exhibit the estimated width of $\sim 8^\circ$ (Rankin 1993),
the central component in the pulse profile of this object
would have to originate from radial distance $\rcore
\simeq67$ km, assuming the uncertain values of 
$\alpha=24^\circ$, $\beta=4^\circ$ and $\thfootp=1$ from Rankin (1990). 
This value is larger than the
error of $\hem$ estimated for this object in Table 1
and should be added to $\hem$ to obtain $\rcone$. 
Had the core component not filled in the open field line region,
the value of $\rcore$ would be even larger.

The RPS method of BCW91 assumes that radio emission at the edge of a profile
originates from \emph{the same} altitudes as the emission 
which determines the center of the position angle curve, in which case
the predicted shift is equal to $4\rdel/\rlc$.
As noted above this assumption,
if not satisfied, would lead to an underestimate
of emission altitudes for the outer edge of a profile.
As shown in Section 3, however, the values of $\redge$
cannot exceed $2\rdel$.

\section{Conclusions}

We have modified the relativistic method of GG2001 of determining the radio
emission geometry within pulsar magnetospheres.
Our modification results in a method of determining the radio emission
altitudes which does not depend on viewing geometry nor does it require
polarization measurements.
According to this method, the altitude of the radio emission
region in units of $\rlc$ is equal to a half of the relativistic phase
shift of a pair of conal components with respect to the core component.
We propose to extend application of the revised method to the outer edge of
radio pulse profiles to identify magnetic field lines associated with
the edge. This may provide a test of the geometric method of
determining $\rem$, based on a measurement of a pulse width at the lowest
intensity level.

As noted by GG2003, the radio emission radii derived with their
original method for the outermost cones
were notoriously larger than the radii derived for the pulse edge
with the geometric method by Kijak \& Gil (1997; 1998). 
Our revision has removed part of this
discrepancy. However, since the geometric method assumes 
that the observed radio emission fills in the open field line region
entirely
($\thfootp = 1$), the emission radii
derived with the geometric method
should be considered as lower limits for $\redge$
(provided the refraction effects can be neglected).
Therefore, other methods of determining $\redge$ for the pulse edge 
should always yield $\redge \ge \rgeo$.

We determined the values of $\thfootp$ for the pulse edge at the 10\% level
using the relativistic method
of Blaskiewicz et al.~(1991). This has revealed that in individual cases
the method often
implies emission from the region of closed magnetic field lines.
BCW91 associated this with the fact that the emission radii $\rdel$
determined with their method may be smaller than $\redge$ because 
inner parts of pulse profiles can originate from lower altitudes.
We have shown that even accounting for this, 
the value of $\redge$ cannot exceed $2\rdel$, 
which is still
not sufficient to solve the problem.
Apparently, other effects, like the 
rotation-induced sweep-back of the
magnetic field lines,
or refraction effects may be responsible for the discrepancy.

\acknowledgments

JD thanks V.~S.~Beskin for comments on pulsar magnetosphere.
We are grateful to J.~Gil and R.~Gangadhara for their valuable comments 
on the manuscript.
We appreciate the comments of the anonymous referee who 
helped us to improve the paper significantly.
This work was performed while JD held a National Research
Council Research Associateship Award at NASA/GSFC.
The work was also supported by the grant 2P03D.004.24 (JD and
BR).  

\appendix

\section{Derivation of equations (\ref{etab}), (\ref{aber}), and
(\ref{thfoot}).} 
\label{appen}

Let the radio waves in the reference frame corotating with the star
be emitted in the direction $\vkp$
at angle $\zp$ with respect to the rotation axis. Because of the
aberration, in the inertial 
observer frame the emission direction will be $\vk \ne \vkp$, and the
observer will have to be located at an angle $\zobs$ 
with respect to $\vec \Omega$ to detect the
radiation (see Fig.~1). In general, $\zobs$ differs from $\zp$
(eg.~Kapoor \& Shukre 1998).
The unit vectors of the emission directions $\vkp$ and $\vk$ 
are related by the aberration formula:
\begin{equation}
\vk = \frac{\vkp + [\gamma + (\gamma-1)(\vbet\cdot\vkp)/\betcor^2]\vbet}
{\gamma(1 + \vbet\cdot\vkp)}
\label{aberr}
\end{equation}
where $\vbet$ is the local corotation velocity in units of the speed of
light $c$, and $\gamma = (1 - \betcor^2)^{-1/2}$ (eg.~Dyks \& Rudak 2003).
Neglecting the terms of the order of $\betcor^2$ and higher, the above
formula reduces to:
\begin{equation}
\vk(1 + \vbet\cdot\vkp) \approx \vkp + \vbet.
\label{aberr2}
\end{equation}
Eq.~(\ref{thfootapp}) implies that
in the small angle approximation (ie.~for emission from within the open
field line region at altitudes $\rem \ll \rlc$) the value of the term
$\vbet\cdot\vkp$ is limited as follows
\begin{equation}
\vbet\cdot\vkp 
\le \betcor\cos\left[\frac{\pi}{2} - \frac{3}{2}(\remp)^{1/2}\right] = 
\betcor\sin[1.5(\remp)^{1/2}] \sim (\remp)^{3/2},
\label{smang}
\end{equation}
where $\remp=\rem/\rlc \ll 1$ is the radial distance of emission. 
Neglecting this term, one obtains
\begin{equation}
\vk \approx \vkp + \vbet
\label{simplest}
\end{equation}
with accuracy of the order of $(\remp)^{1/2}$ 
(ie.~with 10\% error for radio emission at 1\% of $\rlc$; hereafter we 
will use $\remp = 0.01$ in all estimations). Therefore, 
to the first order in $\remp$, 
the aberration angle $\etab$ between $\vk$ and $\vkp$ is approximately
equal to $\betcor$.
By applying the cosine theorem of spherical trigonometry to the
spherical triangle
ABC in Fig.~1:
\begin{equation}
\cos\zobs = \cos\zp\cos\betcor + \sin\zp\sin\betcor\cos\frac{\pi}{2}
\label{cos1}
\end{equation}
and neglecting the $\betcor^2$ terms and higher, 
one finds that
\begin{equation}
\zobs \approx \zp,
\label{zeq}
\end{equation}
ie.~the aberration does not change the colatitude of photon emission
direction measured with respect to $\vec \Omega$ (to the order of $\betcor$).

Application of the sine theorem to the ABC triangle:
\begin{equation}
\frac{\sin\dphab}{\sin\etab} = \frac{\sin(\pi/2)}{\sin\zobs}
\label{cos2}
\end{equation}
gives, for $\dphab \ll 1$, and $\etab \ll 1$, the
relation:
\begin{equation}
\dphab \approx \frac{\etab}{\sin\zobs},
\label{obv}
\end{equation}
used in eq.~(\ref{aber})
(the minus sign in eq.~(\ref{aber})
has been inserted to account for the fact that larger azimuths correspond 
to earlier detection phases).
Eq.~(\ref{simplest}) implies that $\etab \approx \betcor$, ie.:
\begin{equation}
\etab \approx \remp\sin\theta = \remp\sin(\zobs + \delta),
\end{equation}
where $\theta$ is the colatitude of the emission point 
(measured from the rotation axis) and 
$\delta = \theta - \zobs$ so that $|\delta| \approx 3^{-1}|\beta| 
\le 2^{-1}(\remp)^{1/2}\ll 1$,
with $\beta$ being the impact angle. Thus, for $|\delta| \ll 1$ and
$|\delta| \ll \zobs$, one obtains
\begin{equation}
\etab \approx \remp\sin\zobs\left(\cos\delta +
\frac{\sin\delta}{\tan\zobs}\right) \approx \remp\sin\zobs,
\label{etab2}
\end{equation}
which proves eq.~(\ref{etab}). The neglected term
$\sin\delta/\tan\zobs$ becomes large for
the nearly aligned geometry.
In general, ie.~allowing for the maximum possible 
range of impact angles $\beta$ between $\pm1.5(\remp)^{1/2}$,
the resulting error in $\etab$ could be equal to 10\% 
for $\zobs$ as large as $30^\circ$.
In the context of the RPS models discussed in
this paper, however, the angle $\zobs$ corresponding to the 10\% error 
will be much smaller. This is because the line of
sight must traverse nearly through the center of the open field line
region for the core/cone structure to be discernible, ie.~$\beta \sim
\varepsilon 1.5(\remp)^{1/2}$ with the parameter
$\varepsilon$ considerably smaller than 1.
Adopting a reasonable value for $\varepsilon$, eg.~$\varepsilon \sim 0.2$, 
one finds that the error
of $\etab$ exceeds 10\% when $\zobs \le 6^\circ$.

Inserting eq.~(\ref{etab2}) into eq.~(\ref{obv}) gives $\dphab \approx
\remp$ and proves eq.~(\ref{aber}) 
The independence of $\dphab$
on $\zeta$ results directly from the fact that 
the decrease in the aberrational angle $\etab$ for the line of sight
approaching the rotation axis
(eq.~\ref{etab2})
is cancelled out by the ``small circle" effect (eq.~\ref{obv}).
Thus, we find that \emph{as long as the emission direction in the CF
does not deviate much from the meridional plane, 
the aberrational phase shift
does not depend on viewing geometry (ie.~neither on $\alpha$ nor on
$\zobs$)}.
The simplicity of the delay-radius relation derived in BCW91 has the
same origin.

We now turn to the derivation of eq.~(\ref{thfoot}).
Let $\vec \mu$ be the magnetic moment of the dipole, and let $\vkp$ be
the emission direction tangent to the local magnetic field in the CF. 
The equation of the dipolar magnetic field lines is:
$\rem^{-1}\sin^2\theta_{\rm m} = {\rm const}$, where $\theta_{\rm m}$ is the
colatitude of the emission point in the reference frame with $\hat z$
axis along $\vec \mu$.
This equation implies that
\begin{equation}
\thfoot = \arcsin\left[\left(\frac{\rns}{\rem}\sin^2\theta_{\rm
m}\right)^{1/2}\right], 
\label{leq}
\end{equation}
where $\rns$ is the radius of the neutron star.
The angle $\rho = \angle(\vec \mu, \vkp)$ is given by
\begin{equation}
\cos\rho = \frac{\vec \mu \cdot \vkp}{|\vec \mu|} = \frac{
3\cos^2\theta_{\rm m} - 1}{\left(1 + 3\cos^2\theta_{\rm m}\right)^{1/2}}
\label{rho2}
\end{equation}
By calculating $\theta_{\rm m}$ from eq.~(\ref{rho2}) and inserting
into (\ref{leq}) one obtains eq.~(\ref{thfoot}).
The uncertainty of $\rns$, present in eq.~(\ref{leq}), does not affect
the ratio $\thfoot/\thpc$, as long as $\thfoot \ll 1$.
The measurement errors of $\phl$ and $\pht$
result in the error of $\thfoot$
mainly through the inverse square root dependence on $\rem$.
The error of the cones' separation $W$ 
(which enters the formula (\ref{thfoot}) through
eq.~\ref{ro}) is negligible. 
This is because $\dphobs$ (and, thereby, $\rem$) is calculated
as a \emph{small difference} between $\pht$ and $|\phl|$ 
(with $\pht \simeq |\phl|$) whereas $W$ is a sum of these 
(cf.~eqs.~\ref{ph} and \ref{w0} and the comments in Sec.~2).
The dependence $\thfoot \propto \rem^{-1/2}$ implies that the fractional
error of $\thfoot$ ($\sim 13$\%) 
is approximately two times smaller than the fractional error of $\rem$
(cf.~columns 5 and 7 in Table 1).

\clearpage

\begin{deluxetable}{lcccccc}
\tabletypesize{\scriptsize}
\tablecaption{Altitudes $\hem$ of conal radio emission and surface
colatitudes $\thfootp$ 
of magnetic field lines associated with the cones 
derived with the relativistic phase shift method of GG2001
with the modification applied according to eq.~(6) of Section 2.
The numbers are based on cones' shifts
measured by GG2001 and GG2003 and $\alpha$ and $\betmin$ values
from Rankin (1993) (cf.~table 1 in GG2001 and GG2002). \label{tab1}}

\tablewidth{0pt}
\tablehead{
\colhead{Pulsar} &
\colhead{$P$ [s]} &
\colhead{$\nu$ [MHz]} &   
\colhead{Cone\tablenotemark{a}} &
\colhead{$\hem$ [km]} &
\colhead{$\hem$ [\% of $\rlc$]} &
\colhead{$\thfootp$}
}
\startdata
B0329$+$54 & 0.7145 & 325 & 1 & $150\pm080$ & 0.44 & $0.58\pm0.17$\\
           &        & 325 & 2 & $330\pm060$ & 0.96 & $0.57\pm0.06$\\
           &        & 325 & 3 & $600\pm080$ & 1.74 & $0.57\pm0.05$\\
           &        & 325 & 4 & $880\pm240$ & 2.57 & $0.65\pm0.12$\\
           &        &     &   &          &      &          \\
           &        & 606 & 1 & $120\pm070$ & 0.35 & $0.65\pm0.22$\\
           &        & 606 & 2 & $280\pm050$ & 0.83 & $0.58\pm0.06$\\
           &        & 606 & 3 & $460\pm140$ & 1.35 & $0.61\pm0.11$\\
           &        & 606 & 4 & $640\pm180$ & 1.88 & $0.71\pm0.12$\\
           &        &     &   &          &      &          \\
B0450$-$18 & 0.5489 & 318 & 1 & $230\pm25$ & 0.88 & $0.65\pm0.04$\\
           &        &     &   &          &      &          \\
B1237$+$25 & 1.3824 & 318 & 1 & $160\pm40$ & 0.24 & $0.34\pm0.06$\\
           &        & 318 & 2 & $411\pm30$ & 0.62 & $0.43\pm0.02$\\
           &        & 318 & 3 & $540\pm20$ & 0.81 & $0.62\pm0.01$\\
           &        &     &   &          &      &          \\
B1821$+$05 & 0.7529 & 318 & 1 & $230\pm100$ & 0.64 & $0.49\pm0.12$\\
           &        & 318 & 2 & $320\pm090$ & 0.89 & $0.65\pm0.11$\\
           &        & 318 & 3 & $440\pm080$ & 1.23 & $0.72\pm0.08$\\
           &        &     &   &          &      &          \\
B1857$-$26 & 0.6122 & 318 & 1 & $160\pm50$ & 0.55 & $0.72\pm0.12$\\
           &        & 318 & 2 & $350\pm30$ & 1.20 & $0.81\pm0.04$\\
           &        & 318 & 3 & $480\pm80$ & 1.65 & $0.88\pm0.08$\\
           &        &     &   &          &      &          \\
B2045$-$16 & 1.9617 & 328 & 1 & $1000\pm130$ & 1.06 & $0.31\pm0.030$\\
           &        & 328 & 2 & $1790\pm040$ & 1.91 & $0.35\pm0.005$\\
           &        &     &   &           &      &          \\
B2111$+$46 & 1.0147 & 333 & 1 & $800\pm190$ & 1.66 & $0.28\pm0.04$\\
           &        & 333 & 2 & $1230\pm50$ & 2.54 & $0.41\pm0.01$\\
\enddata

\tablenotetext{a}{Cone numbering is the same as in 
GG2001 and GG2003 (ie.~from the innermost cone outwards).}

\tablecomments{The values of $\thfootp$ have been calculated for $\rcore
\ll \hem$.
Had the assumption been not satisfied (as is probably the case at least 
for B0450$-$18), they
should be understood as upper limits.
Errors of $\thfootp$ are substantially underestimated, because
they do not include uncertainty in $\alpha$ and $\betmin$.}

\end{deluxetable}

\begin{deluxetable}{ll}
\tabletypesize{\scriptsize}
\tablecaption{ The applicability of different methods of determining
$\rem$ to different kind of available data.\label{appl}}
\tablewidth{0pt}
\tablehead{
\colhead{Data needed} & \colhead{Analysis method}   
}
\startdata
M or T profile only & this work (eq.~\ref{radius}), OERPS \\
 & \\
profile of any type + position angle curve & geometric, BCW \\
\enddata

\tablecomments{M and T respectively mean `multiple' and `triple' 
pulse profiles according to the classification scheme of Rankin (1983).
OERPS stands for the `outer edge relativistic phase shift method'
(Section \ref{comparison}) and BCW for the delay-radius method of
Blaskiewicz et al.~(1991).
}

\end{deluxetable}

\begin{deluxetable}{lcccc}
\tabletypesize{\scriptsize}
\tablecaption{Surface colatitudes $\thfootp$ of magnetic field lines
associated with the 10\% intensity level, derived with the method of 
BCW91. \label{tab2}}
\tablewidth{0pt}
\tablehead{
\colhead{Pulsar} & \colhead{$\nu$ [GHz]}   & \colhead{$\thfootp$}   &
\colhead{$\pm1\sigma$ error box} &
\colhead{LOC\tablenotemark{a} [$\sigma$]}
}
\startdata
B0301$+$19 & 0.43 & 1.97 & 1.38 -- 2.64 & 1.7\\
           & 1.42 & 3.03 & 1.97 -- 4.65 & 2.1\\
B0525$+$21 & 0.43 & 2.28 & 1.70 -- 3.16 & 2.6\\
           & 1.42 & 1.82 & 1.29 -- 2.89 & 1.8\\
B1913$+$16 & 0.43 & 1.13 & 0.77 -- 1.44 & 0.4\\
           & 1.42 & 0.91 & 0.69 -- 1.12 & 0.4\\
B1914$+$13 & 1.42 & 0.71 & $<0.80$      & 2.3\\
\enddata

\tablenotetext{a}{Level of consistency of $\thfootp$ with 1, in
units of the standard deviation $\sigma$ 
derived from errors of $\rdel$ and $\rgeo$ 
from table 3 in Blaskiewicz et al.~(1991).}
\end{deluxetable}

\begin{figure}
\epsscale{0.7}
\plotone{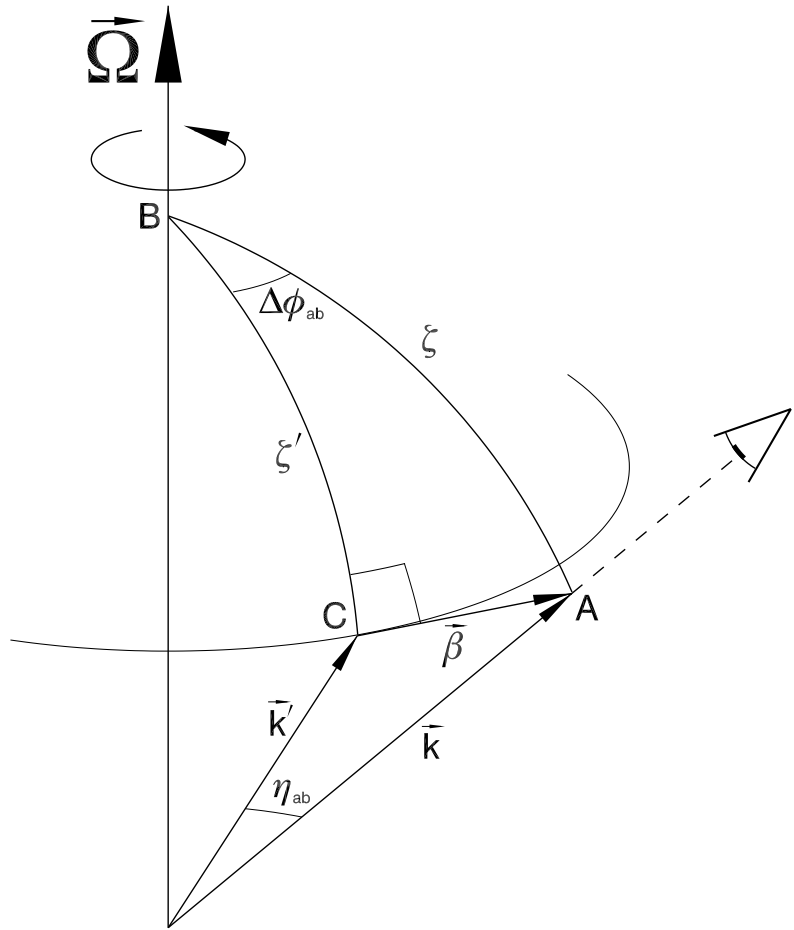}
\caption{Illustration of the geometry considered in the derivation
of eq.~(\ref{aber}) (Appendix A).
\label{spher}}
\end{figure}

\end{document}